# Quantum dynamics of OCS molecule doped in superfluid $^4$He nanodroplet

Ashok K. Jha*, Samrat Dey*^, Yatendra S. Jain*

*Department of Physics, North-Eastern Hill University,
Shillong-794022, Meghalaya, India.

^Assam Don Bosco University, Guwahati-17, India.

In this paper, we explain the change of rotational constant (B) and vibrational frequency shift of $^4$He$_N$-OCS clusters with N by using important inferences of Macro-Orbital theory of superfluidity and several other factors that can change B and vibrational frequency. Consequently, this helps us in understanding the extra ordinary experimental observations that (i) B decreases monotonically from N = 1 to 9 and rises thereafter to follow broad oscillations with maxima at N = 24, 47 and minima at 36 and 62 and (ii) vibrational frequency shift has blue shift for N ~ 1 – 6, with steady red shift thereafter having a change in slope at N ~ 17.

## I. Introduction

In last few decades spectroscopy of molecules embedded in $^4$He droplets has emerged as an interesting area of research for the facts: (i) these droplets can capture virtually any atom or molecule either singly or in the form of clusters, (ii) in these droplet clusters of well-defined size and composition are accessible, (iii) they offer a medium for sensitive detection of absorption, providing highly resolved spectra of doped molecule, (v) they render ultimate spectroscopic matrix to study unperturbed molecular species [1]. While, the first spectroscopic experiment in which single metal atom was embedded in liquid $^4$He was reported in 1985 [2, 3], experiments on molecules were performed in 1992 [4], followed by their high resolution studies in 1995 [5]. First experiment to have a systematic study of ro-vibrational spectra of OCS molecule doped in liquid $^4$He was conducted in 1998 by Grebenev *et.al* [6] with a view to probe the impact of superfluidity of liquid $^4$He. They found that OCS molecule does not rotate freely in liquid $^3$He as well as in $^4$He$_N$-OCS clusters with N < 60; however, in clusters with N > 60 it rotates like a free rotor with slightly increased moment of inertia (*I*), as evident from the sharp rotational peaks. This unique and new observation has been a focus of several research papers on OCS molecule [7-19] and on other molecules [11, 12, 20]. More recently, McKellar *et. al*. [21, 22] have reported free rotation even in clusters of very few $^4$He atoms and discovered a nontrivial dependence of B on N; a similar dependence of vibrational frequency shift has also been observed. To explain these experimental observations (section II), several models (section III) have been used by different research groups but with little success. In this paper we, therefore, use important inferences of Macro-Orbital theory (Section IV) and other significant scientific arguments to explain the said observations. We also give a detailed description of our approach in section V and conclusion of this study in section VI. . We find that our results agree closely with experiments.

## II. Important aspects of experimental spectra

The plot of B vs. N (Fig. 1) illustrates [21, 22] that there is an initial drop in B for N = 1-8, followed by a turn around at N = 9 with a slight kink at N = 14. The B continues to rise between N = 15-24 showing a broad maximum centered at N = 24. B (= 0.085 cm$^{-1}$) at this point lies significantly above its nanodroplet limit (0.073 cm$^{-1}$). The gradual decrease of B towards this limit is modulated by broad oscillations with maxima at N = 24 and 47 and minima at N = 36 and 62. An interesting feature of IR spectrum of $^4$He$_N$-OCS clusters is the observation of vibrational frequency shift (Fig.2) having initial blue shift (for N = 1-5) with a turnaround at N > 5 and a steady red shift thereafter with a change in slope around N = 18 [21, 22].

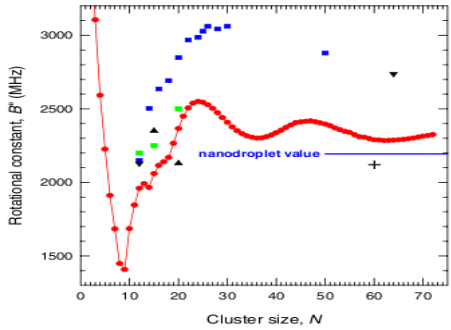 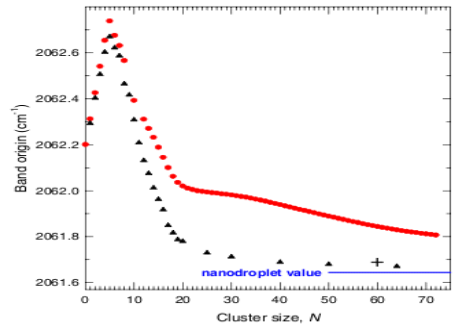

Figure 1: Variation of B with N. Red circles represents experimental values [21, 22]. The observed B for N ≈ 60 clusters [6] within $^3$He nanodroplet is shown as black cross. Theoretical values for N≥12 are shown by black triangles [23, 24], as inverted black triangles [25], as green squares [26], as blue squares [27].

Figure 2: Variation of vibrational band origin with N [21,22]. Red circles represent experimental values. Theoretical values [24] from diffusion Monte Carlo simulation are represented by black triangles. The values observed [6] for N ≈ 60 clusters within $^3$He nanodroplet are shown as black. cross.

### III. Existing theoretical models

Three different theoretical models have been used to explain the experimental observations of different molecules, in general. The first model [5, 10], known as *supermolecule model* (*donut model* in case of OCS molecule), assumes that only specific number of $^4$He atoms rotate rigidly with the molecule and contributes to its *I*. However, this model not only fails to account for the observations for $^4$He$_N$-OCS clusters at quantitative level but also has serious difficulties in explaining experimental observations in case of light rotors like HCN [7]. Second model, known as *two fluid model* [6], tries to relate the free rotation of a molecule with super fluidity of $^4$He. It proposes that the density of $^4$He around the molecule is partitioned into spatially dependent normal fluid and superfluid fractions of which only the former rotates rigidly with the molecule. However, this model could not completely define *spatially dependent normal fluid fraction*. Kwon and Whaley [28] introduced a new concept of non-super fluid density (= total density - super fluid density) that rotates rigidly with the molecule, by assuming that the molecule-$^4$He interaction is sufficiently an-isotropic. Callegari *et.al.* [7, 29, 30] have developed another model, known as *quantum hydrodynamic model*, which adiabatically separates $^4$He motion from molecular rotation; one expects helium density in rotating frame of the molecule to be constant and equal to that around a static molecule [30]. An alternative derivation of quantum hydrodynamic model has been worked put in [31]. As concluded in [17, 32, 33], all these models have some limitations and they have not considered certain realities like: (i) $^4$He atoms around the embedded molecule are localized in a small space (of the order of inter particle distance) where it is impossible to imagine any atom in p = 0 state and this contradicts the conventional picture of superfluidity, (ii) the assumption that the first solvation layer has maximum contribution to normal fluid component, believed to be responsible for the increase in *I* for heavy rotors, does not explain the experimental observations for light rotors like HCN, (iii) the normal fluid density is conventionally associated with inertial mass of quasi-particle excitations [34] but in system like helium droplets such excitations hardly exist. Evidently, a viable model that comprehensively explains different aspects of experimental observations is still awaited.

## IV. Macro-Orbital theory

In Macro-Orbital theory [35], each particle in the system represents a pair of particles moving with equal and opposite momenta (q, -q) at their center of mass which moves with momentum K in the laboratory frame. Below certain temperature say (T) these particles assume a state of (q, -q) bound pairs, named as stationary matter waves pairs (SMW pairs). The λ- transition is found to be a consequence of inter- particle quantum correlations clubbed with zero-point repulsion and inter particle attraction; it is an onset of order – disorder of particles in their Φ-space followed simultaneously by their Bose - Einstein condensation as SMW pairs in a state of q = $q_0$ = π/d and K = 0. Particles at T ≤ $T_λ$ acquire collective binding which locks them at <K> = 0 in momentum space, at <r> = λ/2 = d (average interparticle distance) in real space and at ΔΦ= 2n\ π with (n = 1, 2, 3 ...) in Φ-space. The entire system assumes a kind of collective binding and behaves like a single macroscopic molecule. The collective binding is identified as an energy gap between superfluid and normal fluid states. The fractional density of condensed particles ($n_{K=0}$ (T)) varies smoothly from $n_{K=0}$ ($T_λ$) = 0 to $n_{K=0}$ (0) = 1.0. The λ transition represents the occurrence of twin phenomena of broken gauge symmetry and phase coherence. In variance with conventional belief, the system does not have single particle p = 0 condensate. Instead all atoms in their ground state have identically equal momentum p = h/2d. The particles are localized within a space of inter-particle distance, d and cease to have relative motion, although, they remain free to move in order of locations on a closed path.

## V. Our approach

We understand that the addition of each $^4$He atom in the cluster can change its structure (symmetry, inter-atomic bond lengths and bond angles) which is obviously expected to change its moment of inertia and vibrational frequency shift. However, this simple argument does not account for the experimentally observed decrease in *I* with increasing N. So, we consider that only a few helium atoms in the $^4$He$_N$ -OCS clusters, occupying positions in a plane perpendicular to the molecular axis, rotate rigidly with the molecule and they define the 'first ring'. When this ring gets saturated, $^4$He atoms occupy positions in a 'second ring', in the molecular plane; this provides a sort of equi-potential ring for the angular change in posture of the rest of the cluster (the rotor) and for this reason, $^4$He atoms of this ring do not follow the rotation of the molecule. This consideration finds support from the fact that atoms in different rings in a quantum vortex move with different speed without interfering with each other. Thus addition of each $^4$He atom near this equi-potential ring does not contribute to *I*. On the contrary, with the increase in number of $^4$He atoms in the ring, the order of rotational symmetry increases which decreases *I*, as inferred from analysis of Mathieu's equation [36] for the rotation of a molecular system in a periodic potential that depends on the angular posture of the molecule. Mathieu's equation also suggests that the increase in the height of potential hills (when $V_n$<< 2B) also decreases *I*. However, the presence of $^4$He atom in second ring may change the planar structure of the first ring, increasing the moment of inertia.

The possible arrangement of the $^4$He atoms in the rotor is depicted in Fig. 3, Fig. 4, Fig. 5, and Fig. 6. It is considered that for N = 3 $^4$He atoms orient themselves around the molecule (having the molecular axis along the y-axis), in the first ring (in X-Z) plane and the subsequent $^4$He atoms go to the same ring, contributing to *I* of the rotor, till N = 8 which corresponds to the minimum value of B (Fig.1). For N = 10 onward, B starts to rise and consequently some of the $^4$He atoms are considered to decouple to go to the said equi-potential ring in Y-Z plane; it may be noted that the molecular size of OCS molecule is about 3.32 Å which is not much different from the space size of 3.57 Å occupied by a $^4$He atom in liquid $^4$He, which may easily allow a set of few $^4$He atoms (presumably more than 4) to form a ring in Y-Z plane that does not rotate with the rotation of the rest of the cluster (rotor). Consequently, for N = 10, we consider that 6 $^4$He atom are in the first ring and contributes to *I*, while, 4 $^4$He atoms fall in the second ring, without making any contribution to *I* of the rotor. In case of 11 $^4$He atoms, 5 $^4$He atoms forms the first of ring and 6 $^4$He atoms falls in second ring, making the effective rotor lighter than that of N = 10. This renders higher value of B for N = 11 (Fig.1). Similarly, for N = 12 we have a structure where 4 atoms are in the first ring and 8 atoms are in the second. For N = 13 onwards, the rise in the value of B with N is quite smooth, indicating that the extra $^4$He atoms are not influencing *I* of the rotor significantly and as such, we consider that these extra $^4$He atoms occupy positions near the second ring. When, the $^4$He atoms have finished occupying all the possible sites near these two rings, we consider the formation of a 3-D saturated shell which is presumed to take place around N = 17, indicated by corresponding a change in the slope of dependence of B and vibrational frequency shift on N (Fig.1 and Fig.2). As $^4$He atoms are added to successive shells, $^4$He-OCS distance in the rotor may increase or decrease when different shells reach saturation and thus, we rightly observe the oscillation in B (Fig.1) and corresponding indicators in the shift of $v_1$ frequency with N (Fig.2). Thus, we find that these arguments render a good account of the experimental observations. We have also developed a fortran program [37] for our calculations of $^4He_N$-OCS distance (d) of the rotor that reproduces the experimental B. Our results for $^4$He-OCS distance are depicted in Fig.7. We find that the variation of d is within a very small range (3 - 4 Å). A more quantitative account of these observations with rigorous calculations will be reported later.

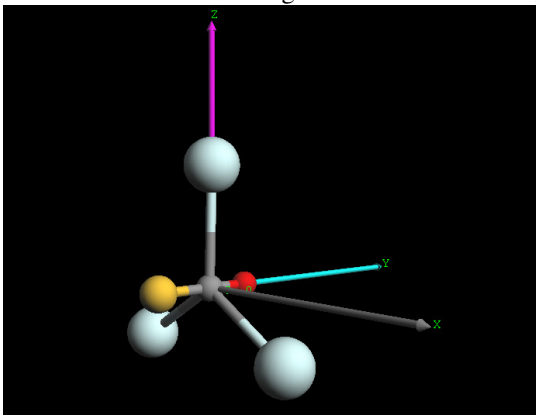

Figure 3: Rotor containing 3 $^4$He atoms around the OCS molecule.

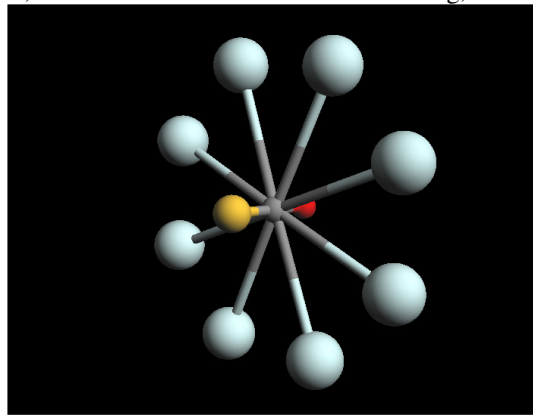

Figure 4: Rotor containing 8 $^4$He atoms around the OCS molecule.

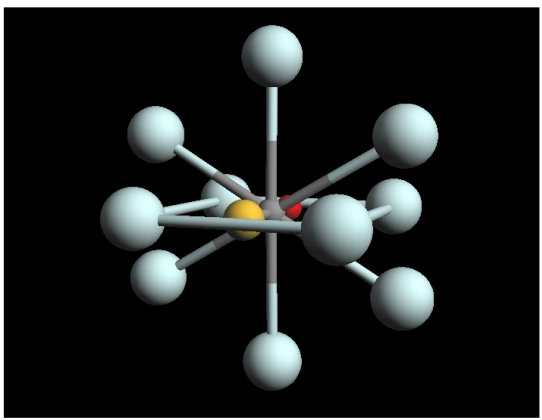

Figure 5: Rotor containing 10 $^4$He atoms around the OCS molecule.

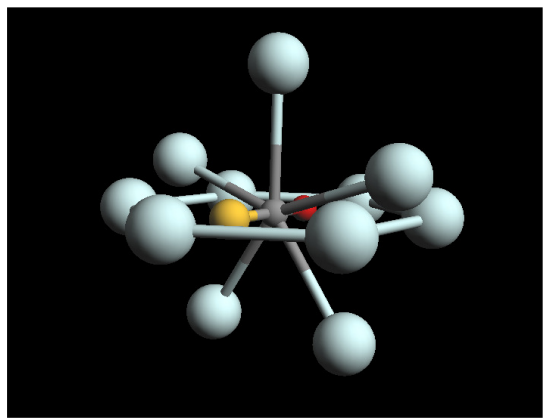

Figure 6: Rotor containing 11 $^4$He atoms round the OCS molecule.

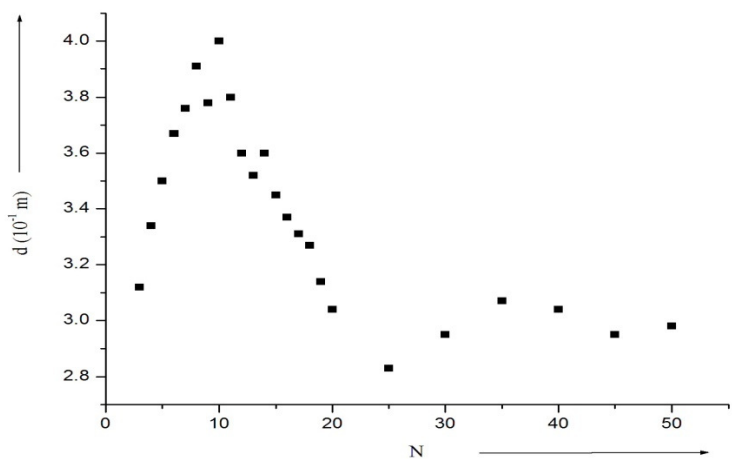

Figure 7: Variation of $^4$He-OCS distance (d) with N.

## VI. Conclusion

Keeping the shortcomings of conventional models in mind, we have developed an effective approach, based on Macro-Orbital theory, to explain the outcomes of spectroscopic experiments on $^4He_N$ - OCS clusters. One may find that our approach provides better understanding of the oscillations in B values and the observed changes in vibrational shift with N. It may be mentioned that our calculations reported here do not consider some significant factors which could also be responsible for the said experimental observations; we shall consider them in our future course of calculations. We hope this will help in finding a comprehensive account of the experiments.